\documentclass{article}

\usepackage{htlab}
\usepackage[numbers]{natbib}

\labname{HiveTraceLab}
\labnamepos{left}

\usepackage[utf8]{inputenc}
\usepackage[T1]{fontenc}
\usepackage{hyperref}
\usepackage{xurl}
\usepackage{tikz}
\usepackage{pgfplots}
\pgfplotsset{compat=1.18}
\usepackage{multirow}
\usepackage{booktabs}
\usepackage{amsfonts}
\usepackage{nicefrac}
\usepackage{microtype}
\usepackage{xcolor}

\title{GLiNER Guard: Unified Encoder Family for Production LLM Safety and Privacy}

\author{
Bogdan Minko\thanks{Corresponding author: \texttt{minkobogdan2001@gmail.com}.}
\quad
Sabrina Sadiekh\thanks{\texttt{sadsobr7@gmail.com}}
\quad
Evgeniy Kokuykin\thanks{\texttt{evgeniy.kokuykin@raftds.com}}
\\
}

\labname{HiveTraceLab}
\lablogo{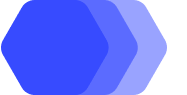}
\lablogowidth{1.3cm}
\labnamepos{left}

\begin{document}
\maketitle

\begin{abstract}
Production LLM systems require both safety moderation and PII detection under strict latency and cost constraints. This creates a trade-off: autoregressive moderators are accurate but expensive, while lightweight encoders are faster but less capable. We present \textbf{GLiNER Guard (GLiGuard)}, a unified encoder that performs safety classification and PII detection in a single forward pass, simplifying safety pipelines. We introduce three variants: compact \textbf{uni-} and \textbf{bi-encoders} (145--147M) for high-throughput serving, and \textbf{GLiGuard Omni} (209M) for stronger moderation quality. Under dynamic batching on a single A100, the compact model reaches 193 requests/sec with P99 latency below 1s, achieving 1.6$\times$ higher throughput than GLiNER2. Omni remains competitive with much larger moderators on public safety benchmarks. We also release \textbf{PII-Bench}, a span-level benchmark for evaluating PII detection in end-to-end pipelines. Overall, encoder-based guardrails offer a practical low-cost alternative for always-on moderation. Models and benchmarks are released on HuggingFace\footnote{Models: \url{https://huggingface.co/collections/hivetrace/gliner-guard-v1}} \footnote{PII-Bench: \url{https://huggingface.co/datasets/hivetrace/pii-bench}}.
\end{abstract}

\section{Introduction}
In production environments, malicious LLM requests, prompt injection, hacking attempts, and user-provided personal data create operational and compliance risks. As a result, requests often need to be screened before reaching downstream systems. Moderation and personal data detection therefore become core components of the first stage of LLM deployment.

Many of the strongest open guardrails are autoregressive models such as LlamaGuard~\cite{llamaguard, llama4guard}, WildGuard~\cite{wildguard}, ShieldGemma~\cite{shieldgemma}, and GPT-OSS-SafeGuard~\cite{gptsafeguard}. These systems can provide high-quality moderation, but their inference cost and latency make always-on deployment expensive at scale. Lightweight encoder models are substantially faster, but existing encoder-based guardrails are typically narrower in scope (e.g., prompt injection only or binary toxicity only) and often less capable. A similar fragmentation exists in privacy pipelines, where production systems frequently rely on a separate NER stack alongside moderation models.
This raises a practical systems question: \textbf{can a compact model provide efficient first-stage protection that combines strong moderation quality and multi-task functionality in a single deployable unit?}

We answer this question with \textbf{GLiNER Guard}, a unified multi-task guardrail built on top of GLiNER2~\citep{gliner2} for safety classification and PII detection in a single forward pass. We study three deployment-oriented variants: (1) a compact uni-encoder, (2) a shared-weight bi-encoder with label caching support, and (3) \textbf{GLiNER Guard Omni}, a larger variant initialized from GLiNER2 Multi to improve transfer and broader generalization. 

Our central contribution is not only adapting GLiNER-style architectures to safety supervision, but showing that a single schema-driven encoder can replace multiple first-stage moderation components under realistic serving constraints. 

Our contributions are:
\begin{itemize}
    \item We introduce GLiNER Guard (GLiGuard), a unified encoder for joint safety classification and PII detection.
    \item We propose compact uni-encoder and shared-weight bi-encoder variants optimized for high-throughput serving.
    \item We present GLiNER Guard Omni, which improves transfer performance while retaining strong moderation quality.
    \item We provide a production-oriented evaluation covering moderation quality, PII detection, generalization, cascading, and serving efficiency.
\end{itemize}
Our results show that compact encoder-based guardrails occupy a practical middle tier between narrow classifiers and large autoregressive moderators, offering strong utility under production latency constraints. This work is a production-oriented technical report. Our focus is practical deployment trade-offs, serving efficiency, and unified first-stage moderation.
\section{Related Work}

\subsection{LLM Safety Guardrails}

\paragraph{Autoregressive guardrails.}
Many of the strongest open safety moderators are autoregressive models trained with safety supervision, including LlamaGuard~\citep{llamaguard}, Llama~4~Guard~\citep{llama4guard}, WildGuard~\citep{wildguard}, ShieldGemma~\citep{shieldgemma}, NemotronGuard~\citep{nemotronguard}, PolyGuard~\citep{polyguard}, and GPT-OSS-SafeGuard~\citep{gptsafeguard}. These systems often perform strongly on response-level moderation, multilingual settings, and longer-context inputs. Their main limitation is serving cost: autoregressive decoding introduces substantially higher latency and inference cost than encoder-only alternatives, making always-on deployment expensive at scale.

\paragraph{Encoder-based guardrails.}
A smaller line of work explores encoder-only models for narrower safety tasks. For example, PromptGuard~2\footnote{\url{https://huggingface.co/meta-llama/Prompt-Guard-2}} focuses on prompt injection and jailbreak detection, Longformer-harmful-ro\footnote{\url{https://huggingface.co/LibrAI/longformer-harmful-ro}} applies Longformer~\citep{beltagy2020longformer} to binary harmful-content classification, and DeBERTa-v3-base-prompt-injection-v2\footnote{\url{https://huggingface.co/ProtectAI/deberta-v3-base-prompt-injection-v2}} uses DeBERTa-v3~\citep{debertav3} for prompt injection detection. Such models are efficient, but they are typically narrow in scope, rely on task-specific fine-tuning, and do not unify moderation with structured extraction tasks such as PII detection.

\subsection{Schema-Driven Information Extraction}
GLiNER~\citep{gliner} introduced a schema-driven alternative to conventional NER: instead of generating labels as text, it scores candidate spans against natural-language entity descriptions, enabling open-label extraction with a compact encoder. GLiClass~\citep{gliclass} extended the same principle to classification tasks. GLiNER2~\citep{gliner2} further unified named entity recognition, relation extraction, and classification within a single encoder architecture conditioned on dynamically defined schemas. However, prior GLiNER-based systems were not designed for safety domains and did not explicitly model adversarial prompts, jailbreak behavior, or moderation taxonomies.

\paragraph{Bi-encoder scalability.}
A bi-encoder variant was later introduced for the original GLiNER by \citet{gliner_bi}. By encoding labels independently and caching their embeddings, it scales more efficiently as label spaces grow. This is particularly attractive for guardrail deployment, where policy schemas may be fixed, large, tenant-specific, or frequently updated. However, prior bi-encoder work focused on single-task NER and relied on two separate encoder towers rather than a shared multitask design.

\subsection{PII Detection}
PII detection is commonly approached through three paradigms. Rule-based systems such as Presidio~\citep{presidio} are highly precise for structured patterns (emails, phone numbers, identifiers) but weaker on context-dependent entities such as names or free-form addresses. Fine-tuned token classifiers based on encoders such as DeBERTa-v3~\citep{debertav3} can achieve strong in-domain performance, but typically depend on labeled data and fixed ontologies. LLM-based extraction approaches such as Llama-3-70B~\citep{llama3} support open-vocabulary extraction, but incur substantially higher inference cost.

Despite strong progress, PII detection is still often deployed as a separate subsystem alongside moderation models. This increases pipeline complexity, maintenance overhead, and latency. A unified model that shares representations across moderation and span extraction offers a simpler systems alternative.

\paragraph{Unified gap and our approach.}
Taken together, prior work reveals three connected limitations: (1) autoregressive guardrails provide strong moderation quality but remain costly for always-on first-stage deployment, (2) existing encoder guardrails are efficient but narrow in scope, and (3) moderation and PII detection are often implemented as separate systems.

GLiNER Guard addresses these limitations with a unified schema-driven encoder family that performs safety classification and PII detection in a single forward pass. We introduce compact uni-/bi-encoder variants for high-throughput serving, a shared-weight bi-encoder with label caching for scalable deployment, and GLiNER Guard Omni for stronger transfer and broader downstream generalization.
\section{Method}

\subsection{Overview}
GLiNER Guard is motivated by a simple observation: safety moderation and PII detection are distinct tasks, but both depend on contextual understanding of the same input text. Built on top of GLiNER2~\cite{gliner2}, the model unifies safety classification and span-level privacy extraction within a single encoder architecture.

The system consists of three components: (1) a shared text encoder, (2) span-scoring heads for extraction tasks, and (3) classification heads for label prediction. Unlike autoregressive guardrails, inference requires no decoding or token generation. Given an input text $x$ and a dynamically defined schema of labels $\{l_i\}_{i=1}^K$, the model predicts classification labels (e.g., safety categories, attack types, intents) and extracts entity spans in a single forward pass.

\subsection{Architecture}

\paragraph{Backbone.}
The compact GLiGuard variants use mmBERT-small~\citep{mmbert} as the backbone, a multilingual adaptation of ModernBERT~\citep{modernbert} with 22 transformer layers, 384-dimensional hidden states, and rotary positional embeddings, supporting 100+ languages. We study three deployment-oriented variants: uni-encoder (147M), bi-encoder (145M) and omni (209M).

\paragraph{Uni-Encoder (147M)}
This variant follows the standard GLiNER2 design: input text and schema labels are concatenated and processed jointly with full bidirectional attention. This provides the richest interaction between text and labels, but requires re-encoding the schema for every request.

\paragraph{Shared-Weight Bi-Encoder (145M)}
Text and labels are encoded separately and projected into a shared embedding space for matching. Unlike prior GLiNER bi-encoders~\citep{gliner_bi}, which use two independent heads for NER only, our design shares a single backbone across both branches and extends the approach to the multi-task GLiNER2 setting. Because label representations are independent of the input text, they can be precomputed and cached when schemas are fixed. This is useful in production deployments with stable or tenant-specific policy taxonomies.

\paragraph{GLiNER Guard Omni (209M)}
The compact uni-/bi-encoder variants are trained from scratch for safety-focused deployment. GLiNER Guard Omni instead starts from GLiNER2 Multi~\citep{gliner2} and is fine-tuned on the same supervision. This preserves more of the base model's general-domain transfer ability while adding guardrail capabilities. Omni is intended for scenarios where broader task coverage or custom-policy support is more important than maximum throughput.

\subsection{Training}

We train all variants on 467,273 multi-task examples, using a 95/5 split between training and held-out validation data. Each sample may contain up to six supervision signals: span extraction, safety classification, adversarial attack detection, harmful-content categorization, intent recognition, and tone classification. This multi-task setup encourages shared representations across related moderation and extraction objectives.

The training mixture is dominated by classification supervision, while span-level NER annotations are substantially less frequent due to the lower availability of high-quality labeled PII data. In total, 108,702 examples contain span extraction labels covering 32 entity types.

For all public datasets, only designated training splits were used. No validation or test examples from evaluation benchmarks were included in training.
\section{Experimental Setup}

Our experiments are designed to answer five practical questions: (q1) how well does the model perform on safety moderation, (q2) what quality--efficiency trade-off does it achieve relative to larger guardrails, (q3) can the same model also support PII detection, (q4) does the Omni variant retain broader transfer beyond safety, and (q5) how efficiently can the system serve requests under realistic load.

\subsection{Evaluation Tracks}

To address these questions, we organize evaluation into three complementary tracks:

\begin{itemize}
    \item \textbf{Safety moderation:} harmful-content detection, adversarial robustness, and multilingual moderation quality.
    \item \textbf{PII detection:} span-level detection of sensitive entities in realistic text.
    \item \textbf{Generalization and serving:} transfer beyond the safety domain and production-oriented inference efficiency.
\end{itemize}

This structure reflects the intended role of a first-stage production guardrail: it must be effective on safety tasks, useful for PII handling, and efficient enough for always-on deployment.

\subsection{Benchmarks}

\paragraph{Safety benchmarks.}
We use three public guardrail benchmarks:

\begin{itemize}
    \item \textbf{Aegis~2.0}~\citep{aegis2}: safety classification for prompts and responses.
    \item \textbf{StrongReject}~\citep{strongreject}: detection of strongly harmful requests.
    \item \textbf{PolyGuard}~\citep{polyguard}: multilingual safety classification with prompt and response splits.
\end{itemize}

For these datasets, we report F1 on the harmful/unsafe class and summarize results with the average score across safety benchmarks (F1$_{\text{avg}}$).

\paragraph{PII benchmarks.}
We evaluate span extraction on two datasets:

\begin{itemize}
    \item \textbf{PII-Bench}: our Russian-language benchmark for span-level PII detection. Because publicly available Russian-language PII benchmarks are limited, we construct a synthetic but human-verified benchmark designed for realistic deployment scenarios. It contains 1,810 examples across 13 entity types and 9 domains. Full benchmark details are provided in Appendix~\ref{sec:app-pii-benchmark}.
    \item \textbf{SPY}~\citep{spy}: a public PII benchmark covering legal and medical domains.
\end{itemize}

\paragraph{Out-of-domain generalization.}
To test whether safety fine-tuning preserves broader capabilities, we evaluate on tasks outside the safety domain:

\begin{itemize}
    \item \textbf{CrossNER}~\citep{crossner}: multi-domain named entity recognition.
    \item \textbf{SST-2}: sentiment classification.
    \item \textbf{Banking77}: intent classification.
\end{itemize}

None of these tasks are included in the safety training data.

\subsection{Baselines}

We compare against three baseline families: guardrail baselines, architectural baselines, and PII baselines.

\paragraph{Guardrail baselines.}
We evaluate autoregressive moderators including LlamaGuard~3~\citep{llamaguard}, Llama~4~Guard~\citep{llama4guard}, WildGuard~\citep{wildguard}, ShieldGemma~\citep{shieldgemma}, NemotronGuardV2~\citep{nemotronguard}, GPT-OSS-SafeGuard~\citep{gptsafeguard}, and YuFeng-XGuard~\citep{yufengxguard}. We also include prior encoder-only guardrails: PromptGuard~2, Longformer-harmful-ro, and DeBERTa-v3-base-prompt-injection-v2.

\paragraph{Architectural baselines.}
To isolate the effect of safety-specific adaptation, we compare against models from the same schema-driven family: GLiNER2~Multi~v1~\citep{gliner2} and GLiClass Instruct Base~v1.0~\citep{gliclass}.

\paragraph{PII baselines.}
For PII detection, we compare against Presidio~\citep{presidio} (rule-based), Llama-3-70B~\citep{llama3} (prompted extraction), and DeBERTa-v3~\citep{debertav3} fine-tuned on SPY.

\subsection{Metrics}

\paragraph{Quality metrics.}
For safety benchmarks, we report F1 on the harmful/unsafe class. For span extraction (PII-Bench and CrossNER), we use strict span matching: a prediction is counted as correct only when the start offset, end offset, and entity type exactly match the reference span. For SPY, we report recall following prior work. For SST-2 and Banking77, we report accuracy. As a parameter-efficiency proxy, we use normalized F1 defined as F1$_{\text{avg}}/\log_2 P$, where $P$ is the number of model parameters.

\paragraph{Serving metrics.}
To evaluate deployment readiness, we report:

\begin{itemize}
    \item \textbf{Throughput (RPS):} sustained requests per second under concurrent load.
    \item \textbf{Latency:} P50, P95, and P99 request latency.
    \item \textbf{Error rate:} fraction of failed or timed-out requests.
\end{itemize}

These metrics complement benchmark quality by capturing practical serving constraints.
\section{Results}

This section is organized around the five practical questions introduced in Section~4. We first evaluate moderation quality and quality--efficiency trade-offs, then test unified PII capability, broader transfer in Omni, and serving efficiency under realistic deployment load.
\subsection{Q1--Q2: Safety Quality and Quality--Efficiency Trade-off}

Table~\ref{tab:benchmarks} compares GLiGuard against autoregressive guardrails, prior encoder baselines, and architecture-matched schema-driven models.

GLiGuard is strongly competitive on safety moderation despite its compact size. \textbf{GLiGuard Omni} achieves the best overall encoder result with 76.9 F1$_{\text{avg}}$, improving over GLiNER2 Multi (66.6) by 10.3 points and GLiClass (64.9) by 12.0 points. On prompt-level moderation, the compact variants remain highly competitive: on Aegis~2.0 prompts, both uni-/bi-encoders reach 80.2 F1, within 1.3 points of WildGuard (81.5), while outperforming LlamaGuard~3 (77.2) and Llama~4 Guard (71.5). On StrongReject, the uni-encoder reaches 98.5 F1 and Omni 99.7, placing both near the top of the comparison set.

GLiGuard also provides a particularly favorable quality--efficiency trade-off. All variants lead the full comparison set on parameter-normalized efficiency (F1$_{\text{avg}}/\log_2 P$), indicating that strong moderation quality is achieved without relying on larger model scale. Figure~\ref{fig:efficiency} visualizes this comparison, with Omni achieving the highest normalized score (2.78).

Table~\ref{tab:latency} provides a complementary single-request view of inference efficiency. Under batch-size-1 evaluation on A100, compact GLiGuard variants remain the fastest encoder models, reaching 54 and 51 requests per second for the bi- and uni-encoder respectively. They also substantially outperform larger autoregressive moderators in latency, for example 0.019 seconds per request for the bi-encoder versus 0.744 for WildGuard.

The largest remaining gap appears in response-level and multilingual moderation, where larger autoregressive models continue to lead. This is consistent with the limitations of compact encoders with fixed context windows and non-autoregressive inference. We address this gap through a cascade design that routes only uncertain cases to a stronger second-stage moderator (Section~\ref{sec:app-cascade}).

\begin{table*}[t]
\centering
\small
\setlength{\tabcolsep}{3pt}
\renewcommand{\arraystretch}{1.03}
\begin{tabular}{llrccccccc}
\toprule
\textbf{Model} & \textbf{Size} & \textbf{Task}
  & \multicolumn{2}{c}{\textbf{Aegis 2.0}}
  & \textbf{StrongReject}
  & \multicolumn{2}{c}{\textbf{PolyGuard}}
  & \textbf{Avg} & \textbf{Avg/$\log_2P$} \\
\cmidrule(lr){4-5}\cmidrule(lr){7-8}
 & & & \textbf{P} & \textbf{R} & & \textbf{P} & \textbf{R} & & \\
\midrule
\multicolumn{10}{c}{\textit{Autoregressive}} \\
\midrule
Llama3Guard       & 8B   & s-cls   & 77.2 & 66.1 & 98.5 & 68.5 & 66.3 & 75.3 & 2.29 \\
Llama4Guard       & 12B  & s-cls   & 71.5 & 64.7 & 95.3 & 62.4 & 54.7 & 69.7 & 2.08 \\
WildGuard         & 7B   & s-cls   & 81.5 & 82.7 & 99.5 & 74.7 & 66.1 & 80.9 & 2.47 \\
ShieldGemma       & 9B   & s-cls   & 79.9 & 74.2 & 90.0 & 47.2 & 41.0 & 66.5 & 2.01 \\
NemotronGuardV2   & 8B   & s-cls   & 86.3 & \textbf{85.4} & 99.5 & 57.5 & 64.4 & 78.6 & 2.39 \\
GPT-OSS-SafeGuard & 20B  & s-cls   & 82.2 & 77.5 & 99.2 & 82.9 & 74.8 & 83.3 & 2.43 \\
YuFeng-XGuard     & 8B   & s-cls   & \textbf{86.4} & 80.4 & \textbf{100.0} & \textbf{85.8} & \textbf{79.2} & \textbf{86.4} & 2.63 \\
\specialrule{1.1pt}{1pt}{1pt}
\multicolumn{10}{c}{\textit{Encoder baselines}} \\
\midrule
Longformer-harmful-ro & 149M & s-cls  & 52.3 & 49.8 & 81.7 & 32.5 & 23.2 & 47.9 & 1.76 \\
PromptGuard $2^*$     & 86M  & s-cls  & 33.7 & 35.0 & 16.4 & 44.3 & 49.3 & 35.7 & 1.35 \\
DeBERTa-v3-base-PI    & 184M & p-inj  & 33.9 & 34.8 & 0.6  & 43.5 & 49.8 & 32.5 & 1.18 \\
\midrule
\multicolumn{10}{c}{\textit{Architecture baselines}} \\
\midrule
GLiClass Base v1.0 & 150M & cls    & 74.1 & 64.5 & 91.3 & 57.6 & 37.1 & 64.9 & 2.39 \\
GLiNER2 Multi v1   & 209M & cls+IE & 71.4 & 75.7 & 95.7 & 60.3 & 29.9 & 66.6 & 2.41 \\
\midrule
\multicolumn{10}{c}{\textit{Ours}} \\
\midrule
GLiGuard bi-enc  & 145M & s-cls+IE & \textbf{80.2} & 74.4          & 97.7          & 69.5          & 49.9          & 74.3          & 2.74 \\
GLiGuard uni-enc & 147M & s-cls+IE & \textbf{80.2} & \textbf{75.7} & 98.5          & 66.2          & 48.5          & 73.8          & 2.72 \\
GLiGuard Omni    & 209M & s-cls+IE & 79.8          & 74.6          & \textbf{99.7} & \textbf{71.7} & \textbf{58.9} & \textbf{76.9} & \textbf{2.78} \\
\bottomrule
\end{tabular}
\caption{Safety moderation results (F1, \%) on Aegis~2.0, StrongReject, and PolyGuard.
Avg is the unweighted mean over the five benchmark columns.
Abbreviations: GLiGuard = GLiNER Guard, P = prompt, R = response.}
\label{tab:benchmarks}
\end{table*}
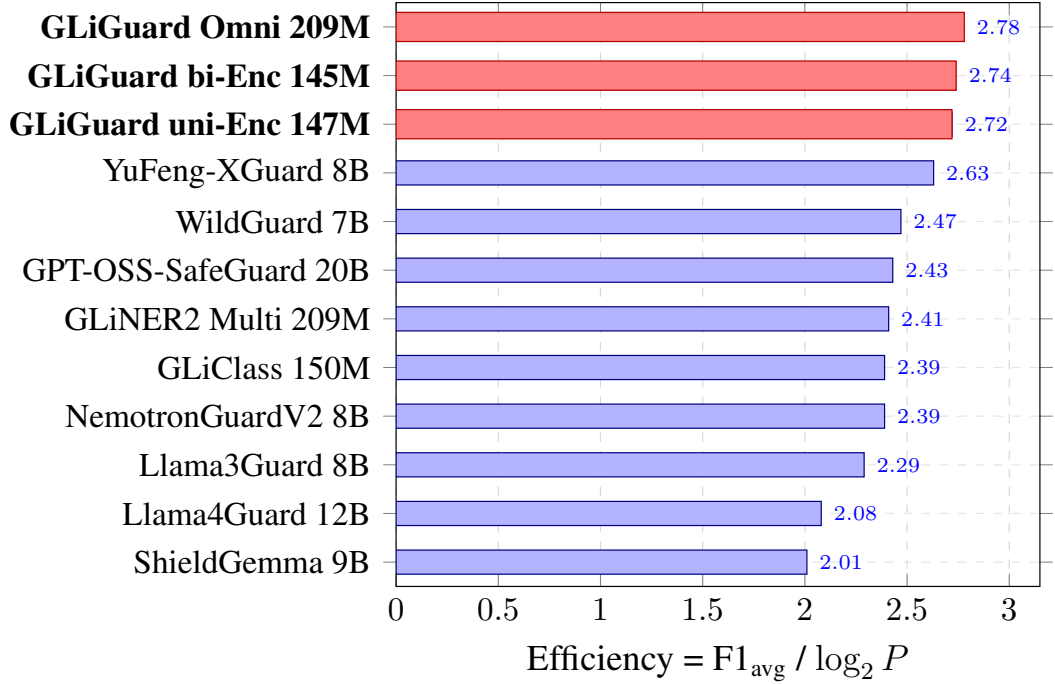
\begin{figure}[t]
  \centering
  \resizebox{\columnwidth}{!}{%
  \begin{tikzpicture}
    \begin{axis}[
      xbar,
      width=0.65\columnwidth,
      height=0.6\columnwidth,
      xlabel={Efficiency = F1$_{\text{avg}}$ / $\log_2 P$},
      xmin=0, xmax=3.15,
      ytick={0,1,2,3,4,5,6,7,8,9,10,11},
      yticklabels={
        ShieldGemma 9B,
        Llama4Guard 12B,
        Llama3Guard 8B,
        NemotronGuardV2 8B,
        GLiClass 150M,
        {GLiNER2 Multi 209M},
        {GPT-OSS-SafeGuard 20B},
        WildGuard 7B,
        {YuFeng-XGuard 8B},
        {\textbf{GLiGuard uni-Enc 147M}},
        {\textbf{GLiGuard bi-Enc 145M}},
        {\textbf{GLiGuard Omni 209M}}
      },
      yticklabel style={font=\small},
      xticklabel style={font=\small},
      nodes near coords,
      nodes near coords style={font=\scriptsize, /pgf/number format/fixed, /pgf/number format/precision=2},
      every node near coord/.append style={anchor=west},
      bar width=8pt,
      grid=major,
      grid style={dashed, gray!20},
      major x grid style={gray!30},
      ymin=-0.5, ymax=11.5,
    ]

    \addplot+[xbar, draw=blue!50!black, fill=blue!30!white] coordinates {
      (2.01, 0)
      (2.08, 1)
      (2.29, 2)
      (2.39, 3)
      (2.39, 4)
      (2.41, 5)
      (2.43, 6)
      (2.47, 7)
      (2.63, 8)
      (2.72, 9)
      (2.74, 10)
      (2.78, 11)
    };

    \fill[red!50!white] (axis cs:0,8.7) rectangle (axis cs:2.72,9.3);
    \draw[red!70!black] (axis cs:0,8.7) rectangle (axis cs:2.72,9.3);
    \fill[red!50!white] (axis cs:0,9.7) rectangle (axis cs:2.74,10.3);
    \draw[red!70!black] (axis cs:0,9.7) rectangle (axis cs:2.74,10.3);
    \fill[red!50!white] (axis cs:0,10.7) rectangle (axis cs:2.78,11.3);
    \draw[red!70!black] (axis cs:0,10.7) rectangle (axis cs:2.78,11.3);

    \end{axis}
  \end{tikzpicture}%
  }
  \caption{Parameter efficiency: F1$_{\text{avg}}$ / $\log_2 P$, where $P$ is the number of parameters. Higher is better. GLiNER Guard achieves the best quality-per-parameter ratio.}
  \label{fig:efficiency}
\end{figure}
\begin{table}[h]
\centering
\small
\begin{tabular}{lrrr}
\toprule
\textbf{Model} & \textbf{Params} & \textbf{Latency$\downarrow$ (s/req)} & \textbf{Throughput$\uparrow$ (req/s)} \\
\midrule
YuFeng-XGuard        & 8B   & 0.051 & 20 \\
WildGuard            & 7B   & 0.744 & 1.3 \\
GLiNER2 Multi        & 209M & 0.021 & 49 \\
\midrule
\multicolumn{4}{c}{\textit{Ours}} \\
\midrule
GLiGuard bi-enc  & 145M & 0.019 & 54 \\
GLiGuard uni-enc & 147M & 0.020 & 51 \\
\bottomrule
\end{tabular}
\caption{Single-request inference speed on A100 80\,GB (batch size 1). Compact GLiNER Guard variants provide the strongest encoder latency while remaining substantially faster than larger autoregressive moderators.}
\label{tab:latency}
\end{table}

Extended serving diagnostics under concurrent load are provided in Appendix~\ref{sec:app-extended-serving}.

\subsection{Q3: Unified Safety and PII Detection}

We next evaluate whether the same deployed model can also support practical PII detection. We report results on \textbf{PII-Bench} in two settings: \emph{model-only} inference and the \emph{full internal NER pipeline}. This distinction is important because the deployed pipeline combines learned span extraction with deterministic pattern-matching rules and post-processing.

In our pipeline, the model is responsible primarily for two context-dependent entity types: \textsc{Name} and \textsc{Address}. Structured entities such as phone numbers, card numbers, tax identifiers, and tokens are detected by rule-based components regardless of the underlying model. For this reason, Table~\ref{tab:pii-name-addr} focuses on \textsc{Name} and \textsc{Address}, which isolate the model's actual contribution. Full per-domain and per-entity results, including rule-based categories, are provided in Appendix~\ref{sec:app-pii}.

The results reveal a clear split between raw extraction quality and final pipeline behavior. On model-only \textsc{Name} extraction, GLiNER2 Multi (85.2 F1) and GLiGuard Omni (83.1) achieve the strongest scores, indicating that broader NER pretraining remains beneficial before post-processing. However, after label mapping and span consolidation, the compact GLiGuard variants achieve the best final pipeline results: the uni-encoder reaches 75.7 F1 on \textsc{Name}, outperforming GLiNER2 Multi (60.6) by 15.1 points, while the bi-encoder achieves the strongest \textsc{Address} result at 68.7 F1, compared with 52.1 for GLiNER2 Multi. These gains are particularly important because they arise exactly on the context-dependent entity types that cannot be handled reliably by simple pattern matching.

A consistent pattern is that raw \textsc{Address} scores are near zero across all models, whereas pipeline scores improve substantially after post-processing. This happens because the benchmark expects one consolidated \textsc{Address} span, while models often predict granular components such as city, street, and unit separately. The pipeline merges these fragments into a single span, making evaluation closer to real redaction behavior.

The \textsc{Name} results show a different trade-off. GLiNER2 Multi and Omni are stronger in raw extraction, but their scores drop after mapping and span consolidation, suggesting that they more often predict fragmented sub-spans whose boundaries do not align with the benchmark's full-name annotations. In contrast, the compact GLiGuard variants produce the strongest final pipeline results.

Overall, these results support the central multitask claim of GLiGuard: a single encoder can provide both safety moderation and useful PII handling within one deployed system, reducing the need for separate moderation and NER stacks.

\begin{table}[!htbp]
  \centering
  \small
  \begin{tabular}{lcccc}
    \toprule
    & \multicolumn{2}{c}{\textbf{NAME F1}} & \multicolumn{2}{c}{\textbf{ADDRESS F1}} \\
    \cmidrule(lr){2-3} \cmidrule(lr){4-5}
    \textbf{Model} & Model & Pipeline & Model & Pipeline \\
    \midrule
    GLiGuard uni-enc  & 74.6 & \textbf{75.7} & 0.5           & 65.9 \\
    GLiGuard bi-enc   & 63.7 & 69.5          & 0.0           & \textbf{68.7} \\
    GLiGuard Omni     & 83.1 & 52.3          & \textbf{6.6}  & 54.6 \\
    GLiNER2 Multi         & \textbf{85.2} & 60.6 & 5.2           & 52.1 \\
    GLiNER2 Large         & 27.8 & 30.3          & 0.0           & 35.9 \\
    \bottomrule
  \end{tabular}
  \caption{PII-Bench: F1\,(\%) on model-dependent entity types. ``Model'' = raw inference; ``Pipeline'' = with span merging and label mapping. GLiNER Guard leads on pipeline \textsc{name}; ADDRESS requires merging for all models.}
  \label{tab:pii-name-addr}
\end{table}

Extended PII breakdowns and SPY results are provided in Appendix~\ref{sec:app-pii}.

\subsection{Q4: Generalization Beyond Safety}

We next test whether safety fine-tuning preserves transfer ability for adjacent tasks and custom policy settings.

The compact uni-/bi-encoders are deliberately specialized for fixed-schema moderation and generalize poorly outside their target domain. On Banking77, for example, they reach only 0.08 and 0.01 accuracy, respectively. This confirms that their strong moderation efficiency comes from deliberate specialization rather than broad transfer capacity.

Omni substantially outperforms the compact variants on both benchmarks, reaching 0.74 accuracy on SST-2 and 0.59 on Banking77. Relative to the original GLiNER2 model, Omni retains much of the base model's zero-shot transfer ability while adding strong moderation performance. This makes Omni the preferable variant when deployments require custom policy categories, broader schemas, or adjacent extraction tasks beyond core safety filtering.

\begin{table}[!htbp]
  \centering
  \small
  \begin{tabular}{lrcc}
    \toprule
    \textbf{Model} & \textbf{Params} & \textbf{SST-2 (Acc)} & \textbf{Banking77 (Acc)} \\
    \midrule
    GPT-4o               & $>$100B & 0.94 & 0.78 \\
    DeBERTa-v3           & 435M    & 0.92 & 0.42 \\
    GLiClass             & 190M    & 0.90 & 0.21 \\
    GLiNER2~\citep{gliner2} & 209M & 0.86 & 0.70 \\
    \midrule
    \multicolumn{4}{c}{\textit{Ours}} \\
    \midrule
     GLiGuard bi-enc  & 145M & 0.50 & 0.01 \\
     GLiGuard uni-enc & 147M & 0.62 & 0.08 \\
     GLiGuard Omni        & 209M & 0.74 & 0.59 \\
    \bottomrule
  \end{tabular}
  \caption{Zero-shot generalization to sentiment (SST-2) and intent detection (Banking77). Accuracy. Uni/bi-encoders collapse on tasks outside their safety training distribution. Omni partially retains generalization from GLiNER2~\citep{gliner2} pretraining, at a cost relative to the base model (0.74 vs.\ 0.86 on SST-2; 0.59 vs.\ 0.70 on Banking77).}
  \label{tab:gen-cls}
\end{table}

Additional CrossNER results are provided in Appendix~\ref{sec:app-generalization}.

\subsection{Q5: Serving Efficiency}
We evaluate serving under dynamic batching and concurrent load on a single A100 80\,GB.
Table~\ref{tab:serving} reports three runtime backends for each model: PyTorch FP16,
ONNX CUDA FP16, and ONNX TensorRT FP16.

GLiGuard achieves the strongest overall serving result, reaching 193.6 requests per
second with 480\,ms P50 latency, 750\,ms P95 latency, and 900\,ms P99 latency under
ONNX TensorRT, with zero errors.

Relative to GLiNER2 under the same ONNX TensorRT backend, GLiGuard improves
throughput by 58\% (193.6 vs.\ 122.6 RPS) while reducing tail latency by 36\% at P99
(900 vs.\ 1400\,ms). Similar gains hold across the PyTorch and ONNX CUDA backends.

We also observe stronger runtime stability: GLiGuard maintains zero errors across
all three backends, whereas GLiNER2 under PyTorch exhibits a 12.95\% error rate under
load.

These results support the intended role of GLiGuard as an always-on first-stage
moderation layer with favorable throughput, latency, and serving robustness.

\begin{table}[h]
  \centering
  \resizebox{\textwidth}{!}{%
  \begin{tabular}{llrrrrr}
    \toprule
    \textbf{Model} & \textbf{Runtime} & \textbf{RPS$\uparrow$} & \textbf{P50 (ms)$\downarrow$} & \textbf{P95 (ms)$\downarrow$} & \textbf{P99 (ms)$\downarrow$} & \textbf{Err\,(\%)$\downarrow$} \\
    \midrule
    \multirow{3}{*}{GLiGuard uni-enc}
      & PyTorch FP16          & 148.2 & 570 & 1500 & 1700 & 0.00 \\
      & ONNX CUDA FP16        & 170.6 & 540 &  870 & 1000 & 0.00 \\
      & ONNX TensorRT FP16    & \textbf{193.6} & \textbf{480} & \textbf{750} & \textbf{900} & 0.00 \\
    \midrule
    \multirow{3}{*}{GLiNER2 Multi}
      & PyTorch FP16          &  83.7 & 1200 & 2000 & 2400 & 12.95 \\
      & ONNX CUDA FP16        &  90.8 & 1000 & 1700 & 2100 &  0.00 \\
      & ONNX TensorRT FP16    & \textbf{122.6} & \textbf{740} & \textbf{1200} & \textbf{1400} &  0.00 \\
    \bottomrule
  \end{tabular}%
  }
  \caption{Serving performance under dynamic batching (LitServe, max batch~64, timeout~50\,ms, NVIDIA A100 80\,GB).
  RPS = requests per second; P50/P95/P99 = end-to-end latency percentiles; Err = HTTP error rate.}
  \label{tab:serving}
\end{table}
Full backend comparisons and batch-size-1 latency are provided in
Appendix~\ref{sec:app-extended-serving}.

Overall, the results provide consistent answers to the five practical questions posed in
Section~4. GLiGuard delivers strong moderation quality with a leading
quality--efficiency trade-off, extends the same deployed model to practical PII
detection, and offers high serving throughput suitable for always-on production use.
Within the model family, the compact uni-/bi-encoder variants are best suited for
cost-efficient first-stage filtering, while Omni trades some efficiency for stronger
transfer, broader schema support, and higher-quality fallback or cascade deployments.
\section{Discussion}

\paragraph{A practical middle tier for guardrails.}
Our results suggest that safety systems need not choose only between small but weak classifiers and large but expensive autoregressive moderators. GLiNER Guard occupies a useful middle tier: compact encoder models that remain competitive on core moderation benchmarks while offering substantially lower serving cost and latency. This makes always-on first-stage filtering practical in settings where LLM moderation would be prohibitively expensive.

\paragraph{Different variants for different deployments.}
The three model variants are complementary rather than strictly ranked. The uni-encoder is the default choice for fixed-schema production moderation, combining strong quality with the highest throughput. The bi-encoder is most attractive when label spaces are large, tenant-specific, or frequently updated, since label embeddings can be cached independently of requests. Omni is the preferred option when deployments require broader transfer, custom schemas, or additional tasks beyond safety moderation, trading some throughput for stronger generalization.

\paragraph{Value of unifying moderation and PII detection.}
A key practical contribution is task unification. Production systems often maintain separate stacks for moderation and PII detection, requiring multiple models, multiple inference passes, and duplicated operational overhead. GLiNER Guard shows that both capabilities can be delivered by a single encoder in one forward pass. The strong PII-Bench results indicate that safety classification and span extraction can share useful representations rather than competing for capacity.

\paragraph{Where larger models still help.}
Large autoregressive guardrails retain advantages on response-level moderation, multilingual transfer, and ambiguous cases requiring longer-context reasoning. We view this not as a failure of encoder-based guardrails, but as evidence for tiered moderation architectures. In such systems, GLiNER Guard handles high-volume routine traffic, while a stronger second-stage model is invoked only for uncertain or structurally difficult inputs.
\section{Limitations and Future Work}

\paragraph{Response-level moderation and context length.}
Response-level moderation remains weaker than the strongest large-model baselines, particularly in multilingual and longer-context settings requiring more advanced reasoning capabilities. The compact context window further constrains robustness on very long inputs.
\paragraph{PII evaluation scope.}
PII-Bench is synthetic and currently limited to Russian-language data. Broader multilingual and real-world privacy evaluation therefore remains unresolved. In addition, some PII categories in the deployed pipeline rely primarily on deterministic rule-based components rather than learned extraction. Consequently, the reported end-to-end PII results should be interpreted as hybrid pipeline performance rather than purely model-based capability.

\paragraph{Comparison and serving constraints.}
Comparisons between encoder-based and autoregressive guardrails are constrained by differing inference paradigms and serving assumptions, which complicates direct comparison. Serving experiments are limited to A100-class hardware and may not directly generalize to CPU-only, edge, or lower-memory deployment environments.

\paragraph{Future work.}
Future work should include stronger calibration analysis, broader robustness evaluation, cost-normalized comparisons under matched deployment constraints, longer-context backbones, improved cascade routing strategies, and broader multilingual supervision.
\section{Conclusion}

We presented \textbf{GLiNER Guard}, a unified encoder-based guardrail that performs safety classification and PII detection in a single forward pass. By combining moderation and structured extraction within one model, it reduces pipeline complexity, lowers serving cost, and simplifies deployment compared with multi-model safety stacks.

Our experiments show that compact encoder guardrails can be both practical and strong. Under realistic production load, the compact variant reaches 193.6 requests per second on a single A100 with 900\,ms P99 latency and zero serving errors, substantially outperforming GLiNER2 Multi in throughput and tail latency under the same runtime. On public safety benchmarks, \textbf{GLiNER Guard Omni} achieves the strongest overall encoder result with 76.9 F1$_{\text{avg}}$, improving over GLiNER2 Multi by 10.3 points while remaining competitive with much larger autoregressive moderators.

The three released variants target complementary deployment needs. The uni-encoder prioritizes maximum throughput for fixed moderation schemas, the bi-encoder enables scalable serving through label caching for large or evolving taxonomies, and Omni extends the framework toward broader zero-shot transfer and multi-purpose use cases. Across variants, we also demonstrate useful zero-shot PII detection and release \textbf{PII-Bench}, a Russian-language benchmark with 1,810 span-annotated examples across 13 entity types and 9 domains.

At the same time, larger autoregressive moderators still hold advantages on response-level moderation, multilingual transfer, and other cases requiring longer-context reasoning. This motivates tiered production architectures in which a fast encoder handles high-volume traffic and stronger LLM moderators are reserved for harder or uncertain requests.

Overall, our results show that unified encoder guardrails are a viable foundation for modern safety systems: fast enough for always-on deployment, flexible enough for multi-task moderation, and strong enough to meaningfully reduce reliance on expensive LLM-only solutions.
\begin{ack}
We sincerely thank Urchade Zaratiana, the creator of GLiNER and GLiNER~2, and Ihor Stepanov, the creator of GLiClass, as well as their teams, whose architectural contributions and open-source work greatly inspired and enabled the development of GLiNER Guard.
\end{ack}

\bibliography{biblio}

\appendix
\appendix




\section{Training Details}
\subsection{Training Hyperparameters}
\label{app:training_hyper}

Table~\ref{tab:hyperparams} reports the optimization settings used for all model variants. The compact uni-/bi-encoder models share the same configuration, while Omni differs in backbone and batch size.

\begin{table}[h]
\centering
\small
\begin{tabular}{lll}
\toprule
\textbf{Parameter} & \textbf{uni/bi-encoder} & \textbf{Omni} \\
\midrule
Backbone            & mmBERT-small & mDeBERTa (GLiNER2-Multi-v1) \\
Max sequence length & 384 & 384 \\
Max span width      & 12 & 8 \\
Epochs              & 3 & 3 \\
Batch size          & 64 & 32 \\
Encoder LR          & $1 \times 10^{-5}$ & $1 \times 10^{-5}$ \\
Task heads LR       & $2 \times 10^{-5}$ & $2 \times 10^{-5}$ \\
Scheduler           & cosine & cosine \\
Optimizer           & AdamW & AdamW \\
Warmup ratio        & 0.1 & 0.1 \\
Weight decay        & 0.01 & 0.01 \\
Max gradient norm   & 10.0 & 10.0 \\
Precision           & bf16 & bf16 \\
\bottomrule
\end{tabular}
\caption{Training hyperparameters. Omni was trained with FlashDeBERTa~\citep{flashdeberta} for memory efficiency and speed.}
\label{tab:hyperparams}
\end{table}

\section{Training Data}
\label{app:training_data}

\paragraph{Data sources and provenance.}
GLiNER Guard is trained on 467,273467,273
467,273 multi-task samples. Each sample carries up to 6 simultaneous tasks: 1 span extraction (NER) and 5 classification tasks. Table~\ref{tab:data-sources} lists the publicly available datasets used to construct the training corpus.
In addition to the public sources, the corpus includes internal data: Russian translations of the WildGuardMix~\cite{wildguardmix} and Aegis~2.0~\cite{aegis2} safety datasets, as well as internal safety data.

\begin{table}[h]
  \centering
  \small
  \begin{tabular}{llll}
    \toprule
    \textbf{Dataset} & \textbf{Domain} & \textbf{Language} & \textbf{License} \\
    \midrule
    WildGuardMix             & Safety      & EN    & ODC-BY \\
    Aegis~2.0                & Safety      & EN    & CC-BY-4.0 \\
    Ru Toxic Texts           & Toxicity    & RU    & MIT \\
    PII NER Corpus Synthetic & PII / NER   & EN    & MIT \\
    Nemotron-PII             & PII / NER   & EN    & CC-BY-4.0 \\
    Synthetic PII Finance    & PII / NER   & Multi & Apache-2.0 \\
    Russian Dialogues        & Dialogues   & RU    & MIT \\
    GrandMaster-PRO-MAX      & General     & RU/EN & Apache-2.0 \\
    \bottomrule
  \end{tabular}
  \caption{Public data sources used in training.}
  \label{tab:data-sources}
\end{table}

\paragraph{Full label distributions.}
Table~\ref{tab:ner-taxonomy} report detailed class frequencies for each training objective.

\begin{table}[p]
  \centering
  \small
  \begin{tabular}{llrr}
    \toprule
    \textbf{Group} & \textbf{Label} & \textbf{Count} & \textbf{\%} \\
    \midrule
    \multirow{5}{*}{Person}
      & \texttt{person} & 52,754 & 13.54 \\
      & \texttt{first\_name} & 13,020 & 3.34 \\
      & \texttt{last\_name} & 7,877 & 2.02 \\
      & \texttt{alias} & 3,493 & 0.90 \\
      & \texttt{title} & 385 & 0.10 \\
    \midrule
    \multirow{10}{*}{Location}
      & \texttt{street} & 26,227 & 6.73 \\
      & \texttt{city} & 24,529 & 6.30 \\
      & \texttt{country} & 12,551 & 3.22 \\
      & \texttt{region} & 8,904 & 2.29 \\
      & \texttt{postal\_code} & 7,954 & 2.04 \\
      & \texttt{unit} & 3,764 & 0.97 \\
      & \texttt{address} & 3,615 & 0.93 \\
      & \texttt{district} & 2,263 & 0.58 \\
      & \texttt{building} & 2,159 & 0.55 \\
      & \texttt{landmark} & 746 & 0.19 \\
    \midrule
    \multirow{5}{*}{Organization}
      & \texttt{company} & 36,365 & 9.33 \\
      & \texttt{product} & 3,196 & 0.82 \\
      & \texttt{government} & 2,105 & 0.54 \\
      & \texttt{education} & 1,193 & 0.31 \\
      & \texttt{media} & 708 & 0.18 \\
    \midrule
    \multirow{4}{*}{Contact}
      & \texttt{email} & 31,024 & 7.96 \\
      & \texttt{phone} & 25,027 & 6.42 \\
      & \texttt{social\_account} & 3,070 & 0.79 \\
      & \texttt{messenger} & 91 & 0.02 \\
    \midrule
    \multirow{3}{*}{Identity}
      & \texttt{national\_id} & 26,782 & 6.87 \\
      & \texttt{document\_id} & 16,912 & 4.34 \\
      & \texttt{passport} & 732 & 0.19 \\
    \midrule
    \multirow{2}{*}{Temporal}
      & \texttt{event\_date} & 31,335 & 8.04 \\
      & \texttt{date\_of\_birth} & 11,571 & 2.97 \\
    \midrule
    \multirow{3}{*}{Financial}
      & \texttt{bank\_account} & 18,845 & 4.84 \\
      & \texttt{card\_number} & 9,304 & 2.39 \\
      & \texttt{crypto\_wallet} & 1,110 & 0.28 \\
    \bottomrule
  \end{tabular}
  \caption{NER entity distribution.}
  \label{tab:ner-taxonomy}
\end{table}


\section{PII-Bench Benchmark Specification}
\label{sec:app-pii-benchmark}

PII-Bench is a Russian-language benchmark for span-level PII detection in realistic production scenarios. It uses explicit character-level offsets, enabling evaluation of complete systems including model predictions, post-processing, and rule-based components. All examples are fully synthetic and do not contain real user data; human verification was used to validate formatting, span boundaries, and domain realism. Two annotators independently reviewed all examples and resolved disagreements through discussion to reach consensus.

Each example is a JSON object containing a text field, a domain label, and a list of entity spans with character-level offsets (start, end) and entity type.

The benchmark covers 13 PII entity types common in Russian-language interactions. Each type has exactly 70 examples in the entity-level split, and every such example contains at least one PII span.

\begin{table}[h]
  \centering
  \small
  \begin{tabular}{llr}
    \toprule
    \textbf{Type} & \textbf{Description} & \textbf{N} \\
    \midrule
    NAME               & Full names (all grammatical cases) & 70 \\
    PHONE\_NUMBER      & Phone numbers & 70 \\
    EMAIL              & Email addresses & 70 \\
    ADDRESS            & Physical addresses & 70 \\
    BANK\_CARD\_NUMBER & Bank card numbers & 70 \\
    CVC                & CVC/CVV codes & 70 \\
    INN                & Taxpayer ID & 70 \\
    KPP                & Tax registration code & 70 \\
    OGRN               & State registration number & 70 \\
    OGRNIP             & Individual entrepreneur reg.\ number & 70 \\
    SNILS              & Social insurance number & 70 \\
    PASSPORT\_NUMBER   & Passport numbers & 70 \\
    TOKEN              & API tokens, recovery keys & 70 \\
    \bottomrule
  \end{tabular}
  \caption{PII-Bench entity types. Each type contains exactly 70 examples with PII spans.}
  \label{tab:app-pii-bench-entities}
\end{table}

Examples are drawn from 9 domains split into two sensitivity levels. \textbf{S} (Sensitive) domains represent settings where real PII is expected and recall is the main priority, since missed entities are more costly than false alarms. \textbf{L} (Low Sensitivity) domains represent more general dialogues where PII is typically absent and false positives are therefore more disruptive to user experience.

\begin{table}[h]
  \centering
  \small
  \begin{tabular}{llrr}
    \toprule
    \textbf{Domain} & \textbf{Description} & \textbf{N} & \textbf{With PII} \\
    \midrule
    S-BANK     & Banking support      & 100 & 65 (65\%) \\
    S-TELECOM  & Telecom support      & 100 & 62 (62\%) \\
    S-DELIVERY & Delivery service     & 100 & 51 (51\%) \\
    S-AUTO     & Auto service         & 100 & 58 (58\%) \\
    S-HR       & HR / recruiting      & 100 & 65 (65\%) \\
    S-RE       & Real estate          & 100 & 66 (66\%) \\
    S-SUPPORT  & General support      & 100 & 55 (55\%) \\
    L-CHAT     & Chats / messengers   & 100 & 50 (50\%) \\
    L-DIALOG   & Multi-turn dialogues & 100 & 50 (50\%) \\
    \bottomrule
  \end{tabular}
  \caption{PII-Bench domains. S-domains prioritize recall; L-domains prioritize low false positive rates.}
  \label{tab:app-pii-bench-domains}
\end{table}

PII-Bench consists of two complementary splits. The \textbf{entity split} groups examples by entity type (NAME, PHONE\_NUMBER, etc.), with each example containing exactly one PII type; it is intended for per-type quality measurement. The \textbf{domain split} groups examples by realistic scenario (banking, telecom, delivery, and so on), with a natural mix of PII and non-PII examples; it is intended for end-to-end pipeline evaluation, including false positive assessment on clean text.

\begin{table}[h]
  \centering
  \small
  \begin{tabular}{lrrr}
    \toprule
    \textbf{Split} & \textbf{Total} & \textbf{With PII} & \textbf{Without PII} \\
    \midrule
    Entity  & 910  & 910 (100\%) & 0 (0\%) \\
    Domain  & 900  & 522 (58\%)  & 378 (42\%) \\
    \midrule
    Total   & 1{,}810 & 1{,}432 (79\%) & 378 (21\%) \\
    \bottomrule
  \end{tabular}
  \caption{PII-Bench sample-level statistics by split.}
  \label{tab:pii-bench-samples}
\end{table}

While 79\% of examples contain PII, the actual character-level PII density is much lower, about 19.6\%, reflecting realistic text distributions in which sensitive spans are short fragments embedded in longer passages.

\begin{table}[h]
  \centering
  \small
  \begin{tabular}{lrrr}
    \toprule
    \textbf{Split} & \textbf{Total chars} & \textbf{PII chars} & \textbf{Clean chars} \\
    \midrule
    Entity  & 65.1K & 18.2K (28.0\%) & 46.9K (72.0\%) \\
    Domain  & 91.7K & 12.6K (13.7\%) & 79.1K (86.3\%) \\
    \midrule
    Total   & 156.8K & 30.8K (19.6\%) & 126.0K (80.4\%) \\
    \bottomrule
  \end{tabular}
  \caption{PII-Bench character-level statistics. PII spans constitute only 19.6\% of total characters.}
  \label{tab:pii-bench-chars}
\end{table}



\section{Extended Serving Diagnostics}
\label{sec:app-extended-serving}

\begin{figure}[h]
  \centering

  \begin{minipage}[t]{0.48\textwidth}
    \centering
    \begin{tikzpicture}
      \begin{axis}[
        ybar, bar width=10pt,
        width=\textwidth, height=6.2cm,
        ylabel={Latency (ms)$\downarrow$},
        ylabel style={font=\small},
        title={GLiNER Guard uni-encoder},
        title style={font=\small\bfseries},
        symbolic x coords={PyTorch FP16, ONNX-CUDA FP16, ONNX-TRT FP16},
        xtick=data,
        xticklabel style={font=\scriptsize, rotate=15, anchor=north east},
        yticklabel style={font=\scriptsize},
        ymin=0, ymax=2900,
        enlarge x limits=0.25,
        nodes near coords,
        nodes near coords style={font=\tiny, rotate=90, anchor=west},
        every node near coord/.append style={
          /pgf/number format/fixed,
          /pgf/number format/precision=0
        },
        legend style={font=\scriptsize, at={(0.99,0.98)}, anchor=north east},
        grid=major,
        grid style={dashed, gray!20},
      ]
        \addplot+ coordinates {(PyTorch FP16,570) (ONNX-CUDA FP16,540) (ONNX-TRT FP16,480)};
        \addplot+ coordinates {(PyTorch FP16,1500) (ONNX-CUDA FP16,870) (ONNX-TRT FP16,750)};
        \addplot+ coordinates {(PyTorch FP16,1700) (ONNX-CUDA FP16,1000) (ONNX-TRT FP16,900)};
        \legend{P50,P95,P99}
      \end{axis}
    \end{tikzpicture}
  \end{minipage}%
  \hfill
  \begin{minipage}[t]{0.48\textwidth}
    \centering
    \begin{tikzpicture}
      \begin{axis}[
        ybar, bar width=10pt,
        width=\textwidth, height=6.2cm,
        ylabel={Latency (ms)$\downarrow$},
        ylabel style={font=\small},
        title={GLiNER2 Multi},
        title style={font=\small\bfseries},
        symbolic x coords={PyTorch FP16, ONNX-CUDA FP16, ONNX-TRT FP16},
        xtick=data,
        xticklabel style={font=\scriptsize, rotate=15, anchor=north east},
        yticklabel style={font=\scriptsize},
        ymin=0, ymax=2900,
        enlarge x limits=0.25,
        nodes near coords,
        nodes near coords style={font=\tiny, rotate=90, anchor=west},
        every node near coord/.append style={
          /pgf/number format/fixed,
          /pgf/number format/precision=0
        },
        legend style={font=\scriptsize, at={(0.99,0.98)}, anchor=north east},
        grid=major,
        grid style={dashed, gray!20},
      ]
        \addplot+ coordinates {(PyTorch FP16,1200) (ONNX-CUDA FP16,1000) (ONNX-TRT FP16,740)};
        \addplot+ coordinates {(PyTorch FP16,2000) (ONNX-CUDA FP16,1700) (ONNX-TRT FP16,1200)};
        \addplot+ coordinates {(PyTorch FP16,2400) (ONNX-CUDA FP16,2100) (ONNX-TRT FP16,1400)};
        \legend{P50,P95,P99}
      \end{axis}
    \end{tikzpicture}
  \end{minipage}

  \vspace{4pt}

  \begin{tikzpicture}
    \begin{axis}[
      xbar,
      bar width=12pt,
      width=\textwidth,
      height=5cm,
      xlabel={Requests per second$\uparrow$},
      xmin=0, xmax=225,
      xticklabel style={font=\scriptsize},
      ytick={0,1,2},
      yticklabels={ONNX-TRT FP16, ONNX-CUDA FP16, PyTorch FP16},
      yticklabel style={font=\scriptsize},
      nodes near coords,
      nodes near coords style={font=\scriptsize},
      every node near coord/.append style={
        anchor=west,
        /pgf/number format/fixed,
        /pgf/number format/precision=1
      },
      legend style={font=\scriptsize, at={(0.99,0.02)}, anchor=south east},
      grid=major,
      grid style={dashed, gray!20},
      ymin=-0.55, ymax=2.55,
      enlarge x limits={upper, value=0.12},
    ]
      \addplot+ coordinates {(148.2,2) (170.6,1) (193.6,0)};
      \addplot+ coordinates {(83.7,2) (90.8,1) (122.6,0)};
      \legend{GLiNER Guard uni-encoder, GLiNER2 Multi}
    \end{axis}
  \end{tikzpicture}

  \caption{Serving performance under dynamic batching (LitServe, max batch~64, timeout~50\,ms, A100 80\,GB). Top: latency percentiles. Bottom: throughput.}
  \label{fig:serving}
\end{figure}
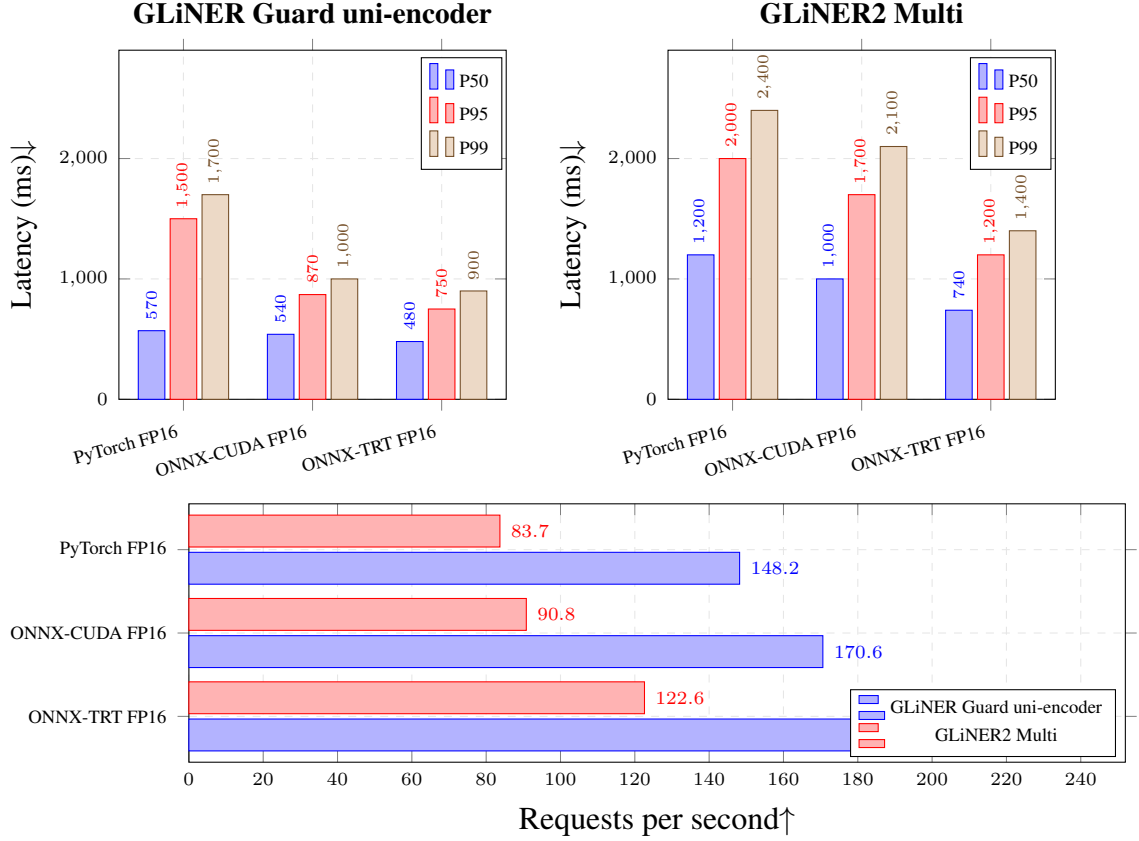

The main paper reports the primary serving results, including best-runtime performance under concurrent load and a single-request latency comparison. This appendix provides the complete backend breakdown across runtimes together with additional serving diagnostics.
We evaluate two serving regimes. Dynamic batching under concurrent load reflects realistic deployment behavior for an always-on first-stage guardrail. Batch-size-1 (see Table~\ref{tab:latency}) latency serves as a hardware-normalized microbenchmark that isolates per-request inference overhead.
Figure~\ref{fig:serving} reports latency percentiles and throughput across three runtimes: PyTorch FP16, ONNX CUDA FP16, and ONNX TensorRT FP16. Across all backends, GLiNER Guard consistently outperforms GLiNER2 Multi in throughput while maintaining lower tail latency. TensorRT provides the strongest performance for both models, but the relative advantage of GLiNER Guard remains stable across runtimes.

\subsection{Extended PII Evaluation}
\label{sec:app-pii}

This section provides the full PII evaluation underlying the summary results reported in the main text. We separate three views of performance: (i) raw model extraction on PII-Bench, (ii) end-to-end pipeline results after deterministic post-processing, and (iii) transfer to the external SPY benchmark.

In the main text, we focus on \textsc{Name} and \textsc{Address} because these are the only entity types whose quality depends materially on the learned model. Structured identifiers such as emails, card numbers, tax IDs, and tokens are handled by deterministic detectors and therefore vary little across model backbones once integrated into the production pipeline.

Table~\ref{tab:pii-model-domain} reports model-only F1 by domain before any span merging or rule-based normalization. GLiNER Guard Omni and GLiNER2 Multi perform strongest in this setting, indicating the benefit of broader NER pretraining for zero-shot span extraction.

\subsubsection{PII-Bench: Raw Model Extraction}

\begin{table}[!htbp]
  \centering
  \resizebox{\textwidth}{!}{%
  \begin{tabular}{lccccc}
    \toprule
    \textbf{Domain} & \textbf{GLiNER Guard uni} & \textbf{GLiNER Guard bi} & \textbf{GLiNER Guard Omni} & \textbf{GLiNER2 Multi} & \textbf{GLiNER2 Large} \\
    \midrule
    L-CHAT     & 75.8 & 86.5 & \textbf{92.6} & 85.5 & 64.3 \\
    L-DIALOG   & 29.9 & \textbf{44.8} & 44.0 & 42.3 & 24.6 \\
    S-AUTO     & 30.0 & 40.5 & 70.6 & \textbf{74.2} & 36.4 \\
    S-BANK     & 16.8 & 23.2 & \textbf{71.6} & 69.4 & 45.8 \\
    S-DELIVERY & 54.2 & 54.6 & 54.1 & \textbf{63.7} & 40.6 \\
    S-HR       & 28.9 & 41.5 & \textbf{82.6} & 78.8 & 52.1 \\
    S-RE       & 21.2 & 27.4 & 64.2 & \textbf{70.1} & 33.8 \\
    S-SUPPORT  & 38.0 & 61.2 & \textbf{81.6} & 74.8 & 63.8 \\
    S-TELECOM  & 24.0 & 33.9 & \textbf{71.1} & 69.9 & 43.9 \\
    \midrule
    \textbf{Avg} & 35.4 & 46.0 & \textbf{70.3} & 69.9 & 45.0 \\
    \bottomrule
  \end{tabular}%
  }
  \caption{PII-Bench: model-only F1\,(\%) by domain.}
  \label{tab:pii-model-domain}
\end{table}

Table~\ref{tab:pii-model-entity} provides the same comparison by entity type. Raw extraction is strongest for structured entities and weaker for entities requiring boundary aggregation, especially \textsc{Address}.

\begin{table}[!htbp]
  \centering
  \resizebox{\textwidth}{!}{%
  \begin{tabular}{lccccc}
    \toprule
    \textbf{Entity} & \textbf{GLiNER Guard uni} & \textbf{GLiNER Guard bi} & \textbf{GLiNER Guard Omni} & \textbf{GLiNER2 Multi} & \textbf{GLiNER2 Large} \\
    \midrule
    EMAIL            & \textbf{100.0} & 97.2  & \textbf{100.0} & \textbf{100.0} & \textbf{100.0} \\
    NAME             & 74.6  & 63.7  & 83.1  & \textbf{85.2}  & 27.8  \\
    PHONE\_NUMBER    & 21.8  & 20.6  & \textbf{87.2}  & 76.0  & 57.5  \\
    BANK\_CARD       & 19.8  & 21.3  & \textbf{69.4}  & 62.7  & 67.3  \\
    CVC              & 21.3  & 0.0   & 40.5  & 27.8  & \textbf{41.1}  \\
    PASSPORT         & 18.8  & 12.8  & \textbf{60.6}  & 51.7  & 16.1  \\
    TOKEN            & 18.6  & 9.7   & \textbf{66.7}  & 33.1  & 14.8  \\
    ADDRESS          & 0.5   & 0.0   & \textbf{6.6}   & 5.2   & 0.0   \\
    INN              & 0.0   & 0.0   & 96.6  & \textbf{97.2}  & 0.0   \\
    KPP              & 5.7   & 0.0   & 81.4  & \textbf{81.9}  & 12.5  \\
    OGRN             & 0.0   & 0.0   & 88.3  & \textbf{99.3}  & 30.4  \\
    OGRNIP           & 19.2  & 0.0   & 90.2  & \textbf{98.6}  & 10.8  \\
    SNILS            & 14.0  & 0.0   & 75.5  & \textbf{75.7}  & 75.0  \\
    \midrule
    \textbf{Avg}     & 24.2  & 17.3  & \textbf{72.8}  & 68.8  & 34.9  \\
    \bottomrule
  \end{tabular}%
  }
  \caption{PII-Bench: model-only F1\,(\%) by entity type.}
  \label{tab:pii-model-entity}
\end{table}

\subsubsection{PII-Bench: End-to-End Pipeline Results}

We next evaluate the full production pipeline, which combines learned extraction with rule-based detectors, label mapping, and span merging. This setting better reflects real deployment behavior than raw model scores alone.

Table~\ref{tab:pii-pipeline-domain} shows pipeline F1 by domain. The compact GLiNER Guard variants achieve the strongest average results, reaching 84.4 and 83.3 F1.

\begin{table}[!htbp]

  \centering

  \resizebox{\textwidth}{!}{%

  \begin{tabular}{lccccc}

    \toprule

    \textbf{Domain} & \textbf{Pipe+Guard uni} & \textbf{Pipe+Guard bi} & \textbf{Pipe+Guard Omni} & \textbf{Pipe+GLiNER2 Multi} & \textbf{Pipe+GLiNER2 Large} \\

    \midrule

    L-CHAT     & \textbf{97.1} & 94.3 & 75.2 & 75.8 & 60.6 \\

    L-DIALOG   & \textbf{76.6} & 75.2 & 63.6 & 47.4 & 25.1 \\

    S-AUTO     & \textbf{74.9} & 74.3 & 73.0 & 71.7 & 58.1 \\

    S-BANK     & \textbf{84.9} & 83.9 & 78.5 & 75.6 & 68.2 \\

    S-DELIVERY & \textbf{86.4} & 83.3 & 75.9 & 77.1 & 61.9 \\

    S-HR       & \textbf{93.0} & 89.4 & 87.7 & 83.8 & 65.9 \\

    S-RE       & 79.8 & \textbf{80.0} & 69.0 & 71.0 & 59.9 \\

    S-SUPPORT  & \textbf{94.4} & 93.8 & 90.7 & 87.2 & 76.8 \\

    S-TELECOM  & 72.3 & \textbf{75.2} & 70.3 & 68.4 & 61.4 \\

    \midrule

    \textbf{Avg} & \textbf{84.4} & 83.3 & 76.0 & 73.1 & 59.8 \\

    \bottomrule

  \end{tabular}%

  }

  \caption{PII-Bench: full pipeline F1 (\%) by domain.}

  \label{tab:pii-pipeline-domain}

\end{table}

Table~\ref{tab:pii-pipeline-entity} separates rule-based and model-dependent entities. As discussed in the main text, the key learned gains come from \textsc{Name} and \textsc{Address}.

\begin{table}[!htbp]

  \centering

  \resizebox{\textwidth}{!}{%

  \begin{tabular}{lccccc}

    \toprule

    \textbf{Entity} & \textbf{Pipe+Guard uni} & \textbf{Pipe+Guard bi} & \textbf{Pipe+Guard Omni} & \textbf{Pipe+GLiNER2 Multi} & \textbf{Pipe+GLiNER2 Large} \\

    \midrule

    \multicolumn{6}{c}{\textit{Rule-based entities (identical across models)}} \\

    \midrule

    EMAIL$^\dagger$         & 100.0 & 100.0 & 100.0 & 100.0 & 100.0 \\

    TOKEN$^\dagger$         & 100.0 & 100.0 & 100.0 & 100.0 & 100.0 \\

    KPP$^\dagger$           & 100.0 & 100.0 & 100.0 & 100.0 & 100.0 \\

    OGRNIP$^\dagger$        & 100.0 & 100.0 & 100.0 & 100.0 & 100.0 \\

    BANK\_CARD$^\dagger$    & 95.5  & 95.5  & 95.5  & 95.5  & 95.5 \\

    OGRN$^\dagger$          & 95.5  & 95.5  & 95.5  & 95.5  & 95.5 \\

    SNILS$^\dagger$         & 95.5  & 95.5  & 95.5  & 95.5  & 95.5 \\

    CVC$^\dagger$           & 94.7  & 94.7  & 94.7  & 94.7  & 94.7 \\

    INN$^\dagger$           & 88.0  & 88.0  & 88.0  & 88.0  & 88.0 \\

    PHONE\_NUMBER$^\dagger$ & 60.4  & 60.4  & 60.4  & 60.4  & 60.4 \\

    PASSPORT$^\dagger$      & 49.6  & 49.6  & 49.6  & 49.6  & 49.6 \\

    \midrule

    \multicolumn{6}{c}{\textit{Model-dependent entities}} \\

    \midrule

    NAME     & \textbf{75.7} & 69.5 & 52.3 & 60.6 & 30.3 \\

    ADDRESS  & 65.9 & \textbf{68.7} & 54.6 & 52.1 & 35.9 \\

    \midrule

    \textbf{Avg} & \textbf{86.2} & 85.9 & 83.5 & 84.0 & 80.4 \\

    \bottomrule

    \multicolumn{6}{l}{\scriptsize $^\dagger$ Detected by deterministic rules; model choice has no effect.}

  \end{tabular}%

  }

  \caption{PII-Bench: full pipeline F1 (\%) by entity type.}

  \label{tab:pii-pipeline-entity}

\end{table}

\subsubsection{External Benchmark: SPY}

Finally, we evaluate transfer to SPY, a public benchmark in legal and medical domains. We report recall only, following prior work, because SPY selectively annotates author-related PII and does not support fair precision comparison.

GLiNER Guard remains strongest on structured entities such as emails, IDs, phone numbers, and addresses, while task-specific DeBERTa-v3 performs best on context-heavy labels such as names and usernames.

\begin{table}[!htbp]
  \centering
  \resizebox{\textwidth}{!}{%
  \begin{tabular}{@{}l ccccccc@{}}
    \toprule
    Model       & \textsf{Presidio} & \textsf{Llama-3-70B} & \textsf{DeBERTa-v3}$^\dagger$ & \textbf{GG (uni)} & \textbf{GG (bi)} & \textbf{GG Omni} \\
    Params      & - & 70\,B & 184\,M & 147\,M & 145\,M & 209\,M \\
    Task        & token cls + rules & autoregressive & token cls & span + text cls & span + text cls & span + text cls \\
    Open-labels & rules only & yes & no & yes & yes & yes \\
    Fine-tuned  & no & no & yes & no & no & yes \\
    \midrule
    \multicolumn{7}{@{}l}{\textbf{(a) Legal Questions}} \\[2pt]
    Name    & 79.4 & 68.9 & \textbf{93.2} & 73.6 & 72.6 & 88.8 \\
    Email   & 91.8 & 88.5 & \textbf{99.1} & 96.0 & 96.0 & 95.9 \\
    User.   & --   & 59.7 & \textbf{98.0} & 40.9 & 43.9 & 43.8 \\
    URL     & 21.3 & 92.5 & \textbf{99.0} & --   & --   & 19.5 \\
    ID      & 34.4 & 62.2 & \textbf{96.6} & 80.7 & 75.8 & 80.9 \\
    Phone   & 68.1 & 92.8 & \textbf{98.7} & 86.8 & 90.4 & 89.4 \\
    Addr.   & --   & 81.3 & 94.5 & 83.4 & 83.8 & \textbf{95.5} \\
    \midrule
    \multicolumn{7}{@{}l}{\textbf{(b) Medical Consultations}} \\[2pt]
    Name    & 80.4 & 62.9 & \textbf{88.7} & 65.9 & 62.3 & 78.2 \\
    Email   & 92.2 & 90.9 & \textbf{99.5} & 94.2 & 94.2 & 93.9 \\
    User.   & --   & 70.4 & \textbf{95.4} & 47.7 & 53.4 & 54.4 \\
    URL     & 19.4 & 91.9 & \textbf{98.9} & --   & --   & 24.1 \\
    ID      & 38.9 & 75.1 & \textbf{98.3} & 89.1 & 84.9 & 86.3 \\
    Phone   & 65.5 & 90.0 & \textbf{96.9} & 86.6 & 89.0 & 87.5 \\
    Addr.   & --   & 90.4 & \textbf{95.1} & 82.2 & 82.7 & 93.7 \\
    \midrule
    \multicolumn{7}{@{}l}{\textbf{Model details}} \\[2pt]
    \multicolumn{7}{@{}p{0.95\linewidth}@{}}{%
      \scriptsize
      \textsf{Presidio}: rule-based NER + regex, zero-shot.\;
      \textsf{Llama-3-70B}: autoregressive prompted NER, zero-shot.\;
      \textsf{DeBERTa-v3}$^\dagger$: token cls, fine-tuned on SPY cross-domain.\;
      \textbf{GG}: GLiNER Guard, zero-shot span extraction + cls, open labels.\;
      \textbf{GG Omni}: fine-tuned from GLiNER~2~Multi on safety-domain data.%
    } \\
    \bottomrule
  \end{tabular}%
  }
  \caption{PII entity recall\;(\%) on SPY. SPY annotates only author-related PII; non-author entities are unlabeled, making precision incomparable across models. ``--'' indicates entity types outside the model's training ontology.}
  \label{tab:spy-recall}
\end{table}




\subsection{Additional Generalization Results}
\label{sec:app-generalization}
We report CrossNER results to complement the SST-2 and Banking77 summary shown in the main text. CrossNER provides a stricter test of open-label transfer because it requires span-level extraction across five unseen domains rather than sentence-level classification alone.

The same pattern observed in the main text becomes even clearer here. The compact uni-/bi-encoder variants, trained for fixed-schema safety deployment, collapse outside their target domain and average only 13.6--14.7 strict F1. This confirms that their strong moderation performance comes from deliberate specialization rather than broad general-purpose transfer.

In contrast, GLiNER Guard Omni retains substantial zero-shot extraction ability after safety fine-tuning, reaching 51.4 average F1 across domains. It performs strongest on Music (58.0), Politics (56.2), and Literature (50.3), showing that the model still transfers to diverse entity schemas beyond safety tasks. Although Omni remains below the original GLiNER2 model (59.0) and GPT-4o (59.9), the gap is moderate relative to the large gains it delivers on moderation benchmarks.

These results reinforce the intended division of roles within the model family: compact variants are optimized for efficient first-stage moderation, whereas Omni offers a better balance between guardrail quality and broader downstream adaptability.

\begin{table}[!htbp]
  \centering
  \resizebox{\textwidth}{!}{
  \begin{tabular}{lrccccccc}
    \toprule
    \textbf{Model} & \textbf{Params} & \textbf{AI} & \textbf{Literature} & \textbf{Music} & \textbf{Politics} & \textbf{Science} & \textbf{Avg} \\
    \midrule
    GPT-4o               & $>$100B & 54.7 & 56.1 & 73.6 & 63.2 & 51.8 & 59.9 \\
    GLiNER-M             & --      & 51.8 & 59.7 & 69.4 & 68.6 & 58.1 & 61.5 \\
    GLiNER2~\citep{gliner2} & 209M & 52.6 & 56.4 & 63.2 & 67.9 & 54.7 & 59.0 \\
    \midrule
    \multicolumn{8}{c}{\textit{Ours}} \\
    \midrule
     GLiNER Guard bi-Encoder  & 145M & 15.3 & 13.5 & 10.1 & 21.2 & 13.4 & 14.7 \\
     GLiNER Guard uni-Encoder & 147M & 14.3 & 11.7 & 10.6 & 19.1 & 12.3 & 13.6 \\
     GLiNER Guard Omni        & 209M & 45.9 & 50.3 & 58.0 & 56.2 & 46.8 & 51.4 \\
    \bottomrule
  \end{tabular}
  }
  \caption{Zero-shot NER strict F1 (\%) on CrossNER across five domains. Uni/bi-encoders are highly specialized for safety tasks, while Omni retains substantially stronger open-domain transfer after safety fine-tuning.}
  \label{tab:gen-crossner}
\end{table}




\subsection{Cascade for Harder Cases}
\label{sec:app-cascade}

The remaining quality gap is concentrated in settings where first-stage encoders are least expected to dominate: long outputs, multilingual inputs, and harder response judgments. We therefore treat this gap as a routing problem rather than a replacement problem.

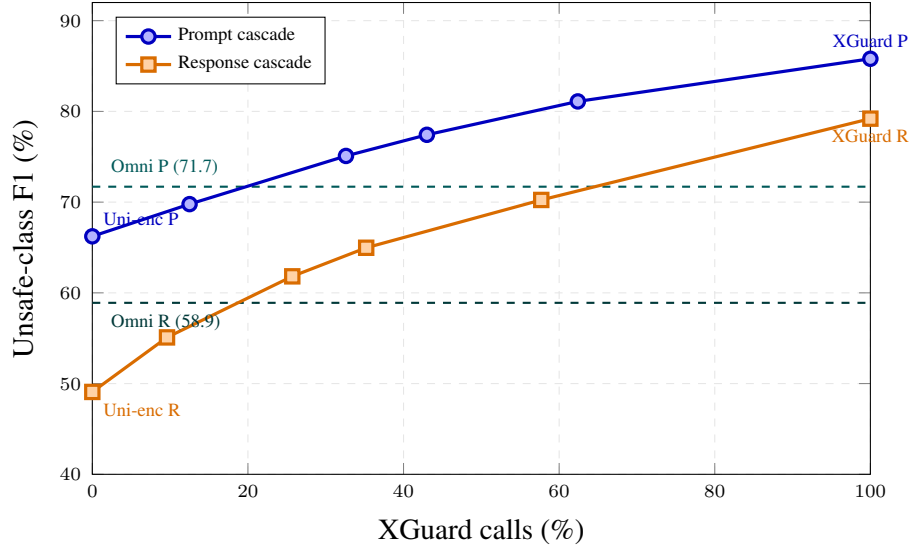
\begin{figure}[t]
  \centering
  \begin{tikzpicture}
    \begin{axis}[
      width=0.85\columnwidth,
      height=0.56\columnwidth,
      xlabel={XGuard calls (\%)},
      ylabel={Unsafe-class F1 (\%)},
      xmin=0, xmax=100,
      ymin=40, ymax=92,
      xtick={0,20,40,60,80,100},
      ytick={40,50,60,70,80,90},
      grid=major, grid style={dashed,gray!20},
      legend style={font=\scriptsize, at={(0.03,0.97)}, anchor=north west,
                    legend cell align=left, legend columns=1},
      xlabel style={font=\small},
      ylabel style={font=\small},
      xticklabel style={font=\scriptsize},
      yticklabel style={font=\scriptsize},
      clip=false,
    ]
    \addplot[teal!70!black, thick, dashed, no markers, forget plot]
      coordinates {(0,71.7)(100,71.7)};
    \addplot[teal!45!black, thick, dashed, no markers, forget plot]
      coordinates {(0,58.9)(100,58.9)};
    \addplot[blue!75!black, very thick,
             mark=*, mark size=2.5pt,
             mark options={fill=blue!30!white, draw=blue!75!black}]
      coordinates {(0,66.23)(12.5,69.78)(32.6,75.09)(43.0,77.42)(62.4,81.10)(100,85.8)};
    \addlegendentry{Prompt cascade}
    \addplot[orange!85!black, very thick,
             mark=square*, mark size=2.5pt,
             mark options={fill=orange!35!white, draw=orange!85!black}]
      coordinates {(0,49.08)(9.6,55.09)(25.7,61.82)(35.2,64.97)(57.7,70.24)(100,79.2)};
    \addlegendentry{Response cascade}
    \node[font=\scriptsize, text=teal!70!black, anchor=south west]
      at (axis cs:1,71.7) {Omni~P (71.7)};
    \node[font=\scriptsize, text=teal!45!black, anchor=north west]
      at (axis cs:1,58.9) {Omni~R (58.9)};
    \node[font=\scriptsize, text=blue!75!black, anchor=south west]
      at (axis cs:0,66.23) {Uni-enc~P};
    \node[font=\scriptsize, text=orange!85!black, anchor=north west]
      at (axis cs:0,49.08) {Uni-enc~R};
    \node[font=\scriptsize, text=blue!75!black, anchor=south]
      at (axis cs:100,85.8) {XGuard~P};
    \node[font=\scriptsize, text=orange!85!black, anchor=north]
      at (axis cs:100,79.2) {XGuard~R};
    \end{axis}
  \end{tikzpicture}
  \caption{Cascade inference on PolyGuard: unsafe-class F1 vs.\ XGuard call rate
  at five GLiNER confidence thresholds ($\tau \in \{0.5,\,0.7,\,0.9,\,0.95,\,0.99\}$).
  \textbf{Solid curves}: cascade (leftmost point = Uni-encoder alone; rightmost = XGuard~8B alone).
  \textbf{Dashed lines}: Omni standalone (no cascade), shown for reference.
  Moving right trades encoder throughput for LLM quality (see Table~\ref{tab:cascade}).}
  \label{fig:cascade}

\end{figure}

Pairing GLiNER Guard with YuFeng-XGuard as a second stage improves moderation quality while routing only a fraction of traffic to the more expensive LLM tier. As shown in Figure~\ref{fig:cascade}, increasing the escalation threshold smoothly trades efficiency for higher quality.

A practical deployment strategy is therefore to use GLiNER Guard as the default moderator and escalate only uncertain or structurally difficult cases.

\begin{table}[t]
  \centering
  \small
  \begin{tabular}{lrccc}
    \toprule
    \textbf{Model} & \textbf{Params} & \textbf{PolyGuard P} & \textbf{PolyGuard R} & \textbf{Type} \\
    \midrule
    Longformer-harmful-ro             & 149M & 32.5 & 23.2 & encoder \\
    GLiNER2 Multi                      & 209M & 60.3 & 29.9 & encoder \\
    Llama4Guard                        & 12B  & 62.4 & 54.7 & autoregressive \\
    GLiGuard uni-Enc \\ (alone)  & 147M & 66.2 & 49.1 & encoder \\
    GLiGuard bi-Enc \\ (alone)   & 145M & 69.5 & 49.9 & encoder \\
    Llama3Guard                        & 8B   & 68.5 & 66.3 & autoregressive \\
   GLiNER Guard Omni \\ (alone)     & 209M & 71.7 & 58.9 & encoder \\
    WildGuard                          & 7B   & 74.7 & 66.1 & autoregressive \\
    \midrule
    Cascade $\tau{=}0.95$ \\ (43\%p / 35\%r calls) & 147M$+$8B & 77.4 & 65.0 & cascade \\
    Cascade $\tau{=}0.99$ \\ (62\%p / 58\%r calls) & 147M$+$8B & 81.1 & 70.2 & cascade \\
    \midrule
    YuFeng-XGuard (alone)              & 8B   & 85.8 & 79.2 & autoregressive \\
    \bottomrule
  \end{tabular}
  \caption{Cascade vs.\ standalone models on PolyGuard (unsafe-class F1, \%).
  P~=~Prompt, R~=~Response. Gray rows are our encoder models; cascade rows combine our encoder with XGuard~8B.
  Cascade operating points interpolate between GLiNER Guard uni-Enc (fast, lower quality) and XGuard (slow, higher quality).}
  \label{tab:cascade}
\end{table}

\end{document}